\documentclass[a4,14pt]{article}
\usepackage[margin=1in]{geometry}
\usepackage{subcaption}
\usepackage{amsmath}
\usepackage{hyperref}
\usepackage{amssymb}
\usepackage{braket}
\usepackage{upgreek}
\usepackage{graphicx}
\usepackage{float}
\usepackage{cite}
\usepackage{esvect}
\begin{document}
\pagenumbering{arabic}
\title{Imprints of vector field non-standard couplings on solar neutrino flux}
\author{H. Yazdani Ahmadabadi\footnote{hossein.yazdani@ut.ac.ir} and H. Mohseni Sadjadi\footnote{mohsenisad@ut.ac.ir}
	\\ {\small Department of Physics, University of Tehran,}
	\\ {\small P. O. B. 14395-547, Tehran 14399-55961, Iran}}
\maketitle
\begin{abstract}
Regarding the proximity of dark energy scale and neutrino mass, one can get new insight into the probe signatures of vector field dark energy component on neutrino flavor conversion.
A vector field is assumed to be conformally coupled to various matter components through a screening paradigm.
We show how the oscillation probabilities evolve differently with a finite-size matter density like the Sun for different values of relevant model parameters.
Considering existing neutrino flavor conversion data, we point out how this non-standard coupling affects neutrino fluxes, and also, we compute the discrepancy in the total flux of solar electron-neutrinos.
\end{abstract}

\section{Introduction}\label{sec1}
An interesting subject in physics nowadays is the problem of dark energy, which is the cause of positive cosmic acceleration.
It seems that there is no obvious evidence for this component in the local observations and experiments \cite{AmendolaBook}.
Among the scalar-tensor dark energy models, screening mechanisms might be described by scalar conformal couplings to different matter components classified as Chameleon \cite{Khoury-PRL,Khoury-PRD,Waterhouse,Tsujikawa} and Symmetron models \cite{Hinterbichler,MohseniSadjadi-SYM1,MohseniSadjadi-SYM2,Honardoost,sad10,BeltranJimenez}. Inspired by this, screening models have been proposed to study the possibility that a field behaves differently in dense and dilute regions.
To detect a scalar or vector field dark energy, which is screened in dense regions, some experiments such as the E$\ddot{o}$t-Wash \cite{Will-GR.EXP,Nobili-WEP} and atom interferometry \cite{Interfero1-Hamilton,Interfero2-Elder,Interfero3-Sabulsky,Interfero4-Lemmel} have been performed.

However, using indirect observational probes can help to assess the possibility of the dark energy existence.
Since the mass scale of neutrinos is similar to the dark energy scale, it is interesting to study the relationship between these two entities \cite{Amendola,Wett1,Wett2,Wett3,Mandal,Salazar-Arias,NSIC2,CarrilloGonzalez,Khalifeh,lam,parv1,parv2}.
Assuming a conformal coupling as an interaction between the massive neutrinos and the scalar field dark energy, the effect of this coupling on neutrino flavor change has been studied in \cite{MohseniSadjadi-Yazdani1}.
The conformal coupling may also be used to relate the quintessence activation to the era when the massive neutrinos became non-relativistic \cite{MohseniSadjadi-Anari1,MohseniSadjadi-Anari2,Sami}.

Analyzing neutrino flavor oscillation data could open a new window to the nature of dark energy.
Neutrino oscillations have been established as the most plausible mechanism behind neutrino flavor transitions \cite{Super-Kamiokande,SNO,KamLAND}, while an additional sub-leading effect caused by neutrino decay has been studied more recently in a two-generation model in \cite{MohseniSadjadi-Yazdani1,MohseniSadjadi-Yazdani3}.
Originating from a damping factor, the net effect of the decay is depletion in the total solar neutrino flux and, consequently, on the electron-neutrino survival probability on the Earth.
The physics beyond the standard model (BSM) may appear as unknown couplings involving neutrinos, usually referred to as non-standard interactions (NSIs) \cite{Miranda,BhupalDev,Ge-Parke}.
Such interactions may modify the neutrino evolution inside matter through an effective Wolfenstein potential \cite{MSWWolf}, induced by vector field interactions with neutrino and matter components \cite{Ge-Parke,Smirnov-WP}.

Neutrinos and the matter components can couple not only to scalar fields but also to vector fields \cite{Ge-Parke,Smirnov-WP}.
Screening models may be extended to vector-tensor models, where the coupling of the vector field to matter makes it vanish in dense regions \cite{BeltranJimenez}.
Like for scalar interactions, the corresponding vector-matter conformal coupling contributes to the modification of several neutrino characteristics, including wavefunction's amplitude and phase, and neutrino effective mass \cite{Smirnov-WP,MohseniSadjadi-Yazdani3,Ge-Parke}.
On the other hand, the scalar field coupling effect is energy-independent, while the vector one scales linearly with neutrino energy.
This leads to significantly different phenomenological consequences in the size of coupling coefficients and, therefore, in solar neutrino oscillations.
Another distinguishing property of vector screening mechanisms with respect to the scalar ones is that the vector case can be considered not only as a manner to screen the fifth force mediated by a vector field, but also as a mechanism to restore Lorentz invariance in dense regions, while being broken in dilute regions \cite{BeltranJimenez}.
For the scalar fields following the screening mechanisms, however, their profile in the presence of matter breaks Lorentz invariance as far as the gradients of the scalar field coupled to the matter energy-momentum tensors, i.e., $\partial_\mu \phi \partial_\nu \phi T^{\mu\nu}_i$, do not vanish \cite{Brax-LV}.

Hence, inspired by \cite{Smirnov-WP,Ge-Parke}, we consider a tree-level neutrino coupling to a vector field that does not belong to the standard model (SM).
We assume that this vector field, which is not detected in usual local tests, obeys a screening mechanism through the conformal coupling to dark and baryonic matters. Indeed, via this mechanism, the vector field is hidden on small scales but may have relevant cosmological impacts \cite{BeltranJimenez}. But as the neutrino and dark energy have similar energy scales, we try to study whether these couplings may have imprints on the neutrino flux, and through this, it might be possible to constrain model parameters.
We present a study of the effective potential induced by a light vector mediator in the context of the screening vector-tensor model.
We consider different couplings as interactions of neutrino and matter components with the dark vector boson, and we compute the Wolfenstein-like potential for a spherically symmetric density profile like the Sun.
We study the new mass and energy dependencies on the coupling constants.
Diagonalizing the neutrino Hamiltonian may introduce the effective matter potential to determine the mass and mixing parameters.

This paper is organized as follows:
In section \ref{sec2}, we review and discuss a vector screening mechanism inside and outside the body, in which the vector field hides on small scales while producing relevant cosmological signatures.
The solution to the Dirac equation in the conformally flat spacetime has been given in section \ref{sec3}, where both the mass and energy of neutrinos obtained new forms.
The vector NSIs with neutrino and matter components, and their phenomenological consequences, are explored in section \ref{sec4}.
This section will also discuss the possibility of neutrino decay and depletion in the total probability.
Furthermore, the vector NSIs might impact the neutrino Hamiltonian, resulting in the effective mass and mixing parameters.
Solar neutrinos can be used to probe for physics BSM that affects neutrino interactions with the vector boson.
Therefore, we restrict our model of neutrino flavor conversion plus decay, including the recent Borexino \cite{Agostini} and SNO+Super-Kamiokande \cite{Zyla} data in section \ref{sec5}.

Throughout the paper, we use units $\hbar=c=1$ and metric signature $(+,-,-,-)$.

\section{Vector screening model}\label{sec2}
Despite the great success of the scalar field dark energy models, one promising approach in the analysis of dark energy is represented by models in which dark energy is described by a vector field, which couples non-minimally to matter components.
In the present model, in which screening occurs due to a spontaneous $O(3)$ symmetry breaking \cite{BeltranJimenez}, the conformal factor depends on the norm of the vector field $A_\mu$.
The general action is
\begin{eqnarray}\label{eqn1}
S= \int d^4x \sqrt{-\text{g}} \left[\frac{M_p^2}{2} \mathcal{R} - \frac{1}{4} F_{\mu\nu} F^{\mu\nu} - \frac{1}{2} \left(\partial_\mu A^\mu\right)^2 - \frac{1}{2} m_A^2 A^2\right] - \int d^4x \mathcal{L}_{\text{m}}\left[\widetilde{\text{g}}_{\mu\nu}, \Psi\right],
\end{eqnarray}
where $\text{g}$ is the determinant of metric $\text{g}_{\mu\nu},$ $\mathcal{R}$ is the Ricci scalar with respect to the metric $\text{g}_{\mu\nu}$, $m_A$ is the mass of the vector field, and $F_{\mu\nu} = \partial_{\mu} A_\nu - \partial_\nu A_\mu$ is the field strength tensor.
The action is decomposed into two parts. The first part includes the exotic vector boson, and the second part describes other ingredients (i.e., matter) conformally coupled to the vector boson through the screening mechanism.
We have included all  matter components $\Psi$ in the Lagrangian $\mathcal{L}_{\text{m}}$.
Recall that a test particle of the matter field $\Psi$ follows the geodesics of the metric $\widetilde{\text{g}}_{\mu\nu}$ related to the metric $\text{g}_{\mu\nu}$ via
\begin{eqnarray}\label{eqn2}
\widetilde{\text{g}}_{\mu\nu}=B^2(A^2) \text{g}_{\mu\nu}.
\end{eqnarray}
Here $A^2 = \text{g}^{\mu\nu} A_\mu A_\nu$ and the conformal coupling function with strength $\beta$ is given by an exponential factor $B(A^2) = \exp (\beta A^2/M_{p}^2)$.
The coupling function $B$ induces an interaction between the vector boson and matter.
We note that the coupled vector boson affects the neutrino equation of motion, and its behavior is almost determined by the matter density, so we encounter an implicit ``matter density'' effect (Wolfenstein-like) on neutrino oscillation through this non-standard interaction.
We use the notation $B^2(A^2)$ because we limit this paper to the transformations that keep the sign of the line element $ds^2$ invariant \cite{Fujii-Maeda}.

By varying the action (\ref{eqn1}), the evolution equation for the massive vector field is given by
\begin{eqnarray}\label{eqn3}
\square A_{\mu} = \left[m_A^2 - \frac{2\beta\rho}{M_p^2} e^{\frac{\beta A^2}{M_p^2}}\right] A_\mu,
\end{eqnarray}
where we have used the rescaled density $\rho \equiv e^{3\beta A^2/M_p^2} \widetilde{\rho}$ for the pressureless matter.
We are interested in the vector field profile for static and spherically symmetric astrophysical sources, such as the Sun.

In the case of a conformal coupling depending on a scalar field $\phi$, one finds the energy-momentum tensor $T_{\mu\nu} = B^2(\phi) \widetilde{T}_{\mu\nu}$\cite{Khoury-PRL}, whereas, in the case of a conformal factor depending on the vector field, we obtain \cite{BeltranJimenez}
\begin{eqnarray}\label{eqn4}
T_{\mu\nu} = B^2(A^2) \left[\widetilde{T}_{\mu\nu} - 2B(A^2)B^\prime(A^2) \widetilde{T}A_{\mu} A_{\nu}\right],
\end{eqnarray}
where the prime denotes differentiation with respect to $A^2$.
In such a case, the additional term appears because the argument of the conformal factor $B(\text{g}^{\mu\nu} A_\mu A_\nu)$ depends on the metric.
This term is necessary only when the matter component is relevant and can be the source of large-scale anisotropic stress \cite{BeltranJimenez}.
Despite the anisotropy of the oscillations, the averaged energy-momentum tensor is isotropic \cite{Cembranos}.

Generally, the field equations (\ref{eqn3}) can be considered a set of four components, coupled to each other with an effective interaction.
This interaction, however, may not be written in terms of a single effective potential, as it is done in the scalar field case.
The medium particles can mostly be considered non-relativistic, hence $\bar{\Psi} \Psi = n_\Psi$ and $\bar{\Psi} \gamma^\mu \Psi = n_\Psi (1,0,0,0)$ \cite{Smirnov-WP}.
Since the number density of solar neutrinos is much smaller than that of electrons or nucleons, the source of the vector field $A_\mu$ is mainly the medium particle and not the neutrinos \cite{Smirnov-WP}.
Thus, we can suppose that the spatial components of $A_\mu$ vanish, i.e., $A_\mu = (A_0,0,0,0)$.
Furthermore, $A_0$ has no time dependence, i.e., $\partial_t A_0 =0$, since the medium particles are at rest.
Therefore, the equation of motion can be approximated by
\begin{eqnarray}\label{eqn5}
\nabla^2 A_0 = - \left[m_A^2 - \frac{2\beta\rho}{M_p^2} e^{\frac{\beta A_0 ^2}{M_p^2}}\right] A_0 = \frac{dV_{\text{eff.}}}{dA_0},
\end{eqnarray}
where the right-hand-side is proportional to the first derivative of an ``effective potential'', which governs the dynamics of $A_0$ and is given by 
\begin{eqnarray}\label{eqn6}
V_{\text{eff.}} := -\frac{1}{2}m_A^2 A_0^2 + \rho~e^{\frac{\beta A_0^2}{M_p^2}},
\end{eqnarray}
which is the sum of its self-interaction term and an exponential term due to the conformal coupling.
Under certain conditions on the self-interacting and coupling parts, this effective potential has two minima, which depend on the local matter density, as does its second derivative at the minimum.
A key point is whether the local matter density $\rho$ allows $A_0$ to experience the $\mathbb{Z}_2$ symmetry breaking, as assumed in the Symmetron scalar field studies \cite{Hinterbichler}.
After the symmetry breaking, the minima of the effective potential are determined using the equation $V_{\text{eff.},A_0}=0$, which leads to
\begin{eqnarray}\label{eqn7}
A_0^{\text{min}} = \pm \sqrt{\frac{M_p^2}{\beta} \ln \left[\frac{m_A^2 M_p^2}{2\beta\rho}\right]}.
\end{eqnarray}
Furthermore, the quantity
\begin{eqnarray}\label{eqn8}
m_{\text{eff.}}^2 = V_{\text{eff.},A_0 A_0}(A_0^{\text{min}})= 2 m_A^2 \ln\left[\frac{m_A^2 M_p^2}{2\beta\rho}\right]
\end{eqnarray}
gives the effective mass of the vector field outside the object (in dilute regions).
Also, note that the vector field's effective mass depends explicitly on the ambient matter density $\rho$.

The effective potential (\ref{eqn6}) exhibits a symmetry breaking mechanism by lowering the density.
At very high densities, the effective potential has only one critical point located at the origin so that the vector field has a vanishing vacuum expectation value.
The effective mass of the fluctuations around $A^{\text{min}}_0 \to 0$ is $m_{\text{eff.}}^2 \simeq 2\beta\rho/M_p^2$.
On the other hand, when the density is low enough, the field starts to run away from the tachyonic critical point at the origin.
In the new vacuum, corresponding to the effective potential after the symmetry breaking, we have two minima with the effective mass given by Eq.(\ref{eqn8}).

The field equation in a static and spherically symmetric profile is given by
\begin{eqnarray}\label{eqn9}
\frac{d^2 A_0}{dr^2} + \frac{2}{r} \frac{dA_0}{dr} = -\left[m_A^2 - \frac{2\beta\rho}{M_p^2} e^{\beta A_0^2 /M_p^2}\right] A_0.
\end{eqnarray}
Given that we are focusing on the Sun, whose Schwarzschild radius is negligible compared to its actual radius, and considering that
$2GM_\odot / R_\odot \sim 10^{-6}$, the effects of spacetime curvature within and outside the Sun are minimal. Therefore, we can safely work within the Minkowskian limit, where $\text{g}_{\mu \nu}=\eta_{\mu \nu}$ \cite{Khoury-PRD}.
Furthermore, this choice for the metric $\text{g}_{\mu\nu}$ is applicable if we ignore the back-reaction of the vector field to the metric itself and, consequently, to the field equation \cite{Khoury-PRD,Khoury-PRL,Waterhouse,Tsujikawa,Hinterbichler}.

According to the symmetry breaking condition, there is a $(\beta,m_A)$ region generated by the vector-matter conformal coupling and allowed by the present bounds:

$\bullet$ For $\rho \gg \frac{m_A^2 M_p^2}{2\beta}$, since the objects of interest, i.e., the Sun and Earth, are much denser than the current critical density, the symmetry is restored and the vector field is screened. In such a case, the effective potential can be approximated such that its first derivative is given by $V_{\text{eff.},A_0} \simeq \frac{2\beta \rho(r) A_0}{M_p^2}$.

\begin{figure}[H]
	\centering
	\includegraphics[scale=0.7]{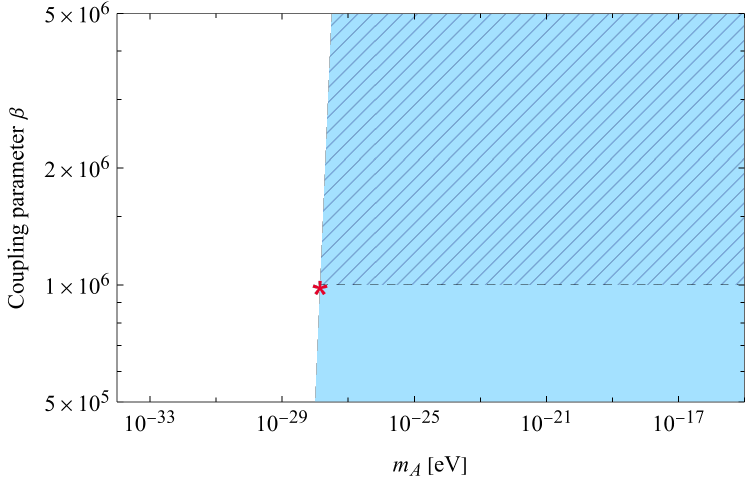}
	\caption{\footnotesize{The plot illustrates the region allowed by $\rho_0 \lesssim \frac{m_A^2 M_p^2}{2\beta}$ implying the condition of symmetry breaking (the blue-hatched region).
		The coordinates of asterisk correspond to $(\beta,m_A) = (10^6,10^{-28} \text{eV})$.}
	}
	\label{fig1}
\end{figure}

$\bullet$ For $\rho_0 \lesssim \frac{m_A^2 M_p^2}{2\beta}$, to find the allowed region in the parameter space, we note that this condition guarantees the symmetry breaking, i.e., the blue-hatched region depicted in Fig. \ref{fig1}.
A wide range of the coupling parameter in which the symmetry is broken outside the object and restored inside is $\beta \geq 10^6$ so that the weak equivalence principle violations and E$\ddot{o}$t-Wash-like experiments can be easily satisfied \cite{BeltranJimenez}.
Thus, the allowed value of the vector field's mass is $m_A \gtrsim 10^{-28}$ eV corresponding to $\beta \geq 10^6$.
In order to solve the field equation, it is instructive to consider the field profile sourced by a spherical object of radius $R_\odot$ embedded in a medium of the Minkowski background density $\rho_0$.
So, using a background-object split, we may expand the vector field around its background value, i.e., $A_0 \equiv \delta A_0 + A_0^{\text{min}}$, where $\delta A_0$ is the field sourced by the spherical object, and $A_0^{\text{min}}$ is the uniform background value (see Eq.(\ref{eqn7})).
The equation for the interior case is then given by
\begin{eqnarray}\label{eqn10}
\frac{d^2 \delta A_0}{dR^2} + \frac{2}{R} \frac{d\delta A_0}{dR} = m_{\text{in}}^2 R_{\odot}^2(A_0^{\text{min}} + \delta A_0), & (R<1)
\end{eqnarray}
where $m_{\text{in}}$ is the effective vector field's mass inside the Sun.

Outside the body, we assume that we are in the symmetry breaking phase.
In such a case, the above equation for the perturbation of the field is given by
\begin{eqnarray}\label{eqn11}
\frac{d^2 \delta A_0}{dR^2} + \frac{2}{R} \frac{d\delta A_0}{dR} = m_{\text{out}}^2 R_{\odot}^2 \delta A_0, & (R>1)
\end{eqnarray}
where $m_{\text{out}}$ is the vector field's effective mass outside the Sun, depending on the background density $\rho_0$.
We have written relations in terms of the dimensionless fractional radius $R \equiv r/R_{\odot}$.

\begin{figure}[H]
	\centering
	\includegraphics[scale=0.61]{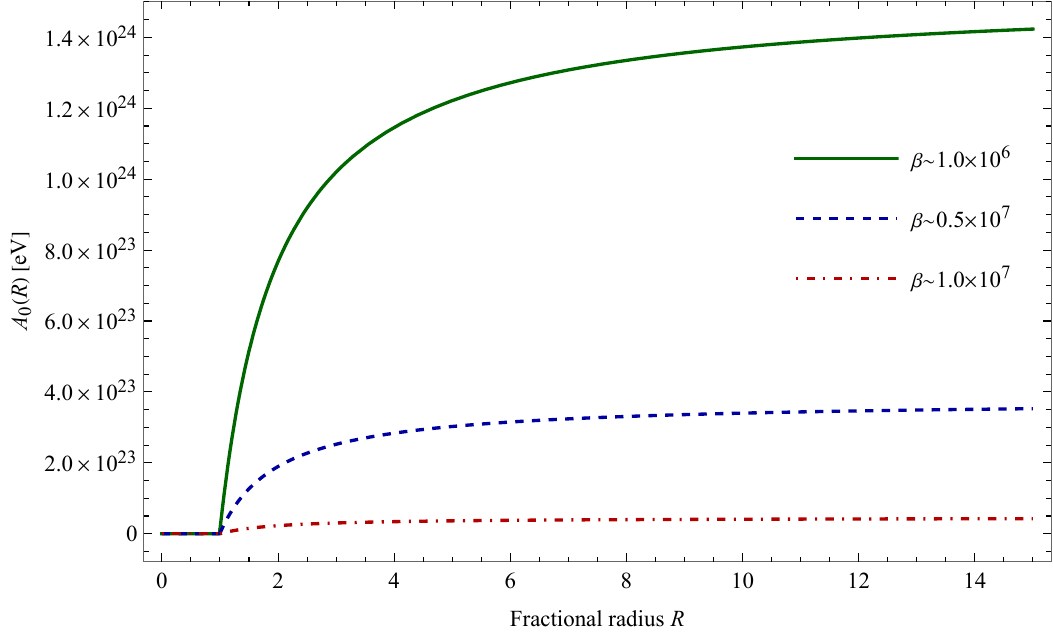}
	\caption{\footnotesize{Shown is the vector field plot with respect to the fractional radius.
			We have tuned different values of $\beta$ for the coupling strengths between vector and matter fields.}
}
	\label{fig2}
\end{figure}

\begin{figure}[H]
	\centering
	\includegraphics[scale=0.45]{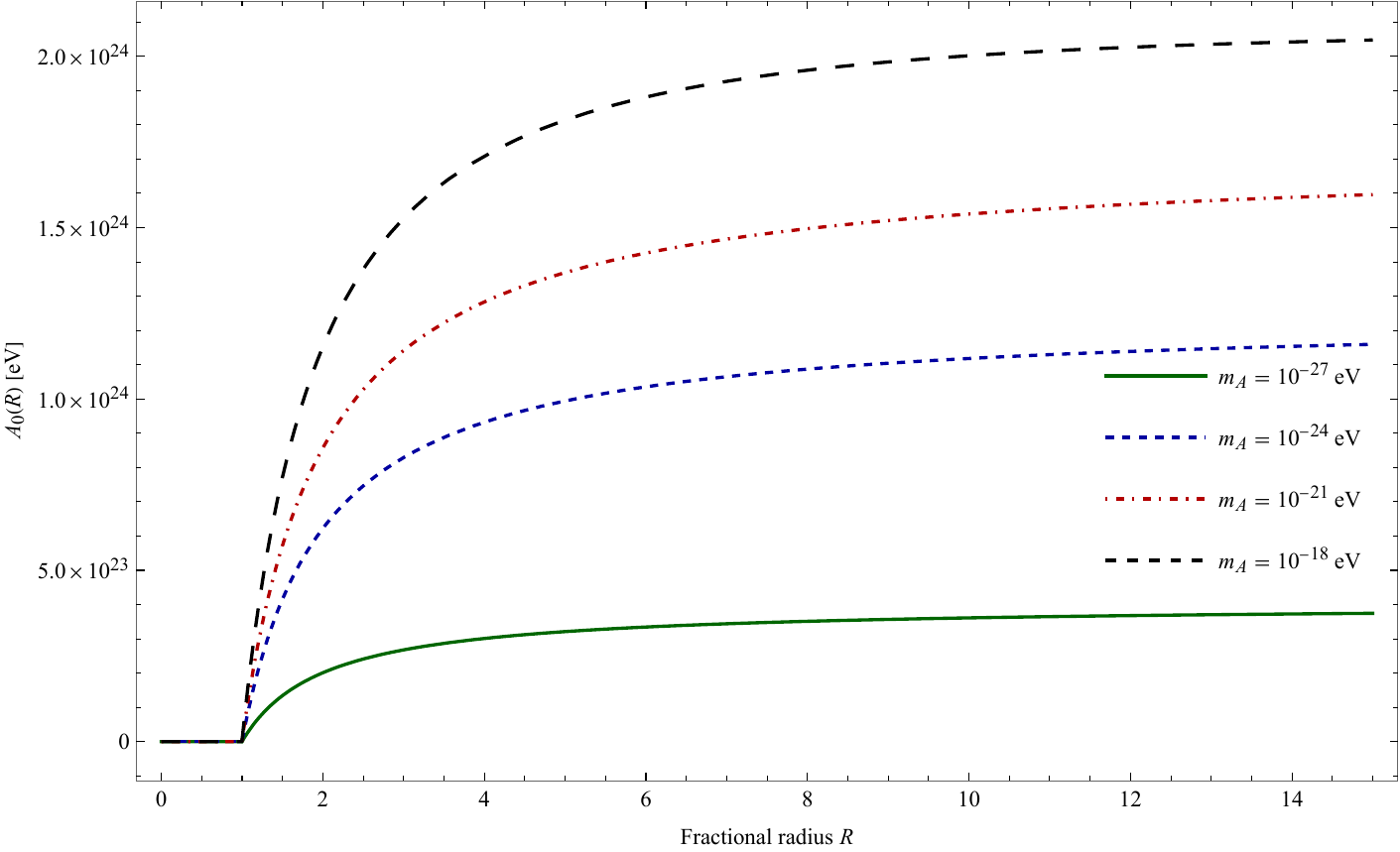}
	\caption{\footnotesize{Evolution of the vector field as a function of the solar fractional radius.
			In this figure, we have used the coupling strength $\beta = 1.0 \times 10^7$, and different curves correspond to the various vector field masses $m_A$.}
	}
	\label{fig3}
\end{figure}

The obtained equation looks similar to those in \cite{MohseniSadjadi-Khosravi,MohseniSadjadi-Yazdani2}, so we can proceed to obtain the solutions inside and outside the object of slowly varying density $\rho_{\odot}$.
We use the following boundary conditions:
\begin{eqnarray}\label{eqn12}
\begin{split}
&\frac{d\delta A_0}{dr} \longrightarrow 0 ~~~ \text{at} ~~ r \to 0 \\&
\delta A_0 \longrightarrow 0 ~~~ \text{at} ~~ r \to \infty
\end{split}
\end{eqnarray}
to obtain a regular solution to these equations in a spherically symmetric static background.
The analytical solutions to the vector field equations (\ref{eqn10}) and (\ref{eqn11}) are
\begin{eqnarray}\label{eqn13}
\delta A_{0}^{\text{in}}(R) = - A_0^{\text{min}} +  \frac{C_{1}}{m_{\text{in}} R_{\odot} R} \sinh \left[m_{\text{in}} R_{\odot} R\right], & (R<1)
\end{eqnarray}
and
\begin{eqnarray}\label{eqn14}
\delta A_{0}^{\text{out}}(R) = C_{2} \frac{1}{R} e^{-m_{\text{out}} R_{\odot} R}. & (R>1)
\end{eqnarray}
To specify the constants $C_{1}$ and $C_{2}$, we should use the continuity of the field and its first derivative at $R = 1$, which leads us to
\begin{eqnarray}\label{eqn15}
C_{1} = \frac{A^{\text{min}}_0 m_{\text{in}}\left(1 + m_{\text{out}} R_{\odot}\right)}{m_{\text{out}} \sinh(m_{\text{in}} R_{\odot}) + m_{\text{in}} \cosh(m_{\text{in}} R_{\odot})},
\end{eqnarray}
and
\begin{eqnarray}\label{eqn16}
C_{2} = - e^{m_{\text{out}} R_{\odot}} \frac{A_0^{\text{min}}}{R_{\odot}} \frac{m_{\text{in}} R_{\odot} \cosh(m_{\text{in}} R_{\odot}) - \sinh(m_{\text{in}} R_{\odot})}{m_{\text{in}} \cosh(m_{\text{in}} R_{\odot}) + m_{\text{out}} \sinh(m_{\text{in}} R_{\odot})}.
\end{eqnarray}
In figure \ref{fig2}, we show the profile corresponding to the above solutions for different vector-matter coupling parameters $\beta$, where we see the analog of the thin-shell effect and how the field profile goes to zero very rapidly inside the Sun.
It is evident that, while inside the Sun, the field value is negligible, the field takes non-zero asymptotic values outside the Sun.
For stronger couplings, the field obtains fewer values, and tends to its asymptotic values at distances closer to the Sun's surface.
Also, we have depicted the vector field profile with respect to the fractional radius $R$ in figure \ref{fig3}. Different curves describe the field $A_0$ corresponding to different values of the relevant parameter $m_A$, which demonstrate the asymptotic behavior outside the Sun. Higher values of mass $m_A$ give larger values of the vector field.
Furthermore, the vector field reaches its asymptotic value for a smaller $m_A$ (green solid curve).

\section{Neutrino flavor change with vector NSI}\label{sec3}
As mentioned, the vector field's behavior is determined by the local matter density in the environment.
A non-standard coupling of neutrinos or other matter components with a massive vector field as a candidate for dark energy is motivated by the tantalizing closeness in scale between neutrino rest mass and the scale of dark energy ($\sim$ meV).
Hence, the ambient matter density might implicitly affect the neutrino flavor transition.
The neutrino sector in the action is given by:
\begin{eqnarray}\label{eqn17}
\widetilde{S}_\nu = \int d^4x \sqrt{-\widetilde{\text{g}}}~\bar{\upnu}(x) \left[i \widetilde{\gamma}^\mu \widetilde{D}_\mu  - m  - g \widetilde{\gamma}^\mu A_\mu\right]\upnu(x),
\end{eqnarray}
where $\widetilde{\text{g}}$ is the determinant of metric $\widetilde{\text{g}}_{\mu\nu}$, and $m$ is the neutrino mass.
The quantity $g$ is defined as the coupling coefficient between neutrino only and the dark vector component, while the parameter $\beta$ has been introduced as vector-matter (except neutrinos) conformal coupling.
We shall suggest a new possibility, the so-called conformal transformation, to govern the procedure, simplifying the derivation of expressions related to the applications of the present model.
Under the conformal transformation (\ref{eqn2}), the gamma matrices and also the covariant derivative are given by \cite{MohseniSadjadi-Yazdani1}
\begin{eqnarray}\label{eqn18}
\widetilde{\gamma}^\mu = B^{-1} \gamma^\mu,
\end{eqnarray}
and
\begin{eqnarray}\label{eqn19}
\widetilde{\gamma}^\mu \widetilde{D}_\mu = B^{-1} \gamma^\mu D_\mu + \frac{3}{2} B^{-2} B_{,\mu} \gamma^\mu.
\end{eqnarray}
Here, all quantities with a tilde correspond to the original metric $\widetilde{\text{g}}_{\mu\nu}$.
By interpreting the conformal coupling (\ref{eqn2}) as an interaction of matter components and the vector field \cite{MohseniSadjadi-Yazdani1}, its effect on neutrino oscillation may be studied by recasting the equation of motion as
\begin{eqnarray}\label{eqn20}
\gamma^\mu (p_\mu + g A_\mu) \upnu^\prime + m^\prime \upnu^\prime = 0
\end{eqnarray}
in the transformed frame with metric $\text{g}_{\mu\nu}$, where $\upnu^\prime \equiv B^{\frac{3}{2}} \upnu$ is the rescaled wavefunction (interpreted as damped neutrino oscillation due to energy exchange with the exotic field $A_\mu$ \cite{MohseniSadjadi-Yazdani1}), and $m^\prime \equiv m B$ is the modified neutrino mass.
So, we encounter an ordinary neutrino oscillation equation (we have taken $\text{g}_{\mu\nu} \approx \eta_{\mu\nu}$ in Eq.(\ref{eqn2})), with an extra coupling coming from $\gamma^\mu A_\mu \upnu^\prime$ (see Eq.(\ref{eqn20})).
This modifies the wavefunction's phase as
\begin{eqnarray}\label{eqn21}
\int p_\mu dx^\mu \to \int \left(p_\mu + gA_\mu - \frac{3}{2} i \partial_\mu \ln B\right) dx^\mu
\end{eqnarray}
in the original frame.
Therefore, with $A_\mu = (A_0,0,0,0)$, the oscillation phase difference for ultra-relativistic neutrinos (i.e., for $t \simeq z$) becomes
\begin{eqnarray}\label{eqn22}
\int_{z_0}^{z} \left[(E_\nu + gA_0) - p_z + \frac{3}{2} i \partial_z \ln B\right] d\text{z},
\end{eqnarray}
where the real parts denote the actual oscillation phase, showing itself in the observations as the oscillatory term, and the imaginary part implies a damping behavior caused by the vector-matter non-standard coupling.
We may analyze these two terms separately:
\paragraph{$\bullet$ Real part}
In the limit $m^{\prime 2} \ll p^2$, and $\gamma^\mu \partial_\mu m^\prime \ll p^2$, we have $E_\nu + gA_0 = p_z \left(1+ \frac{m^{\prime 2}}{2 p_z^2}\right)$.
Therefore, by replacing $p_\mu$ with $p_\mu + gA_\mu$, the ``real'' phase difference for two mass eigenstates, due to $\gamma^\mu A_\mu \upnu^\prime$, acquires the additional potential term
\begin{eqnarray}\label{eqn23}
\int_{z_0}^{z} \frac{\Delta m^{\prime 2}_{ij}}{2 (E_\nu + gA_0)} d\text{z}.
\end{eqnarray}
Hence, the vector field affects the neutrino wavefunction's phase through the Wolfenstein-like potential $V_W \equiv g A_0$, coming from the vector-neutrino NSI \cite{Smirnov-WP}, as well as through the conformal coupling employed in the screening mechanism, which induces an implicit matter density effect following a spontaneous symmetry breaking (see section \ref{sec2}).
\paragraph{$\bullet$ Imaginary part}
Furthermore, by the imaginary part of the phase difference (\ref{eqn22}), we can define a ``damping factor'' caused by the vector-matter conformal coupling as follows:
\begin{eqnarray}\label{eqn24}
\mathcal{D}(z) \equiv \left[B(z) B^{-1}(z_0)\right]^{-3}.
\end{eqnarray}
As a result, non-standard interactions of neutrinos can have effects on the flavor transition.
In the present model, vector interactions not only introduce a new non-standard Wolfenstein-like potential added to neutrino energy, i.e., $E_\nu \to E_\nu + g A_0$, but also affect the neutrino's effective mass (see below Eq.(\ref{eqn20})).
On the other hand, since the scalar interaction cannot be converted into a vector current, light scalar fields, e.g., Chameleon and Symmetron \cite{MohseniSadjadi-Yazdani1,MohseniSadjadi-Yazdani3}, do not give rise to Wolfenstein-like potential.
Instead, they only contribute as a correction to the neutrino mass term.

Collecting all together, the wavefunction of neutrinos is given by
\begin{eqnarray}\label{eqn25}
\begin{split}
&\upnu_i(t,z) = \sqrt{\mathcal{D}(z)} e^{i\varphi_i(t,z)} \upnu_i(t_0,z_0)\\&
~~~~~~~~~ \equiv \mathcal{F}_i(t,z) \upnu_i(t_0,z_0).
\end{split}
\end{eqnarray}
In fact, this damping signature is obtained as a minor perturbation that affects directly the amplitude of various neutrino flavor conversion probabilities.
Damping signatures can be induced by a class of new physics models, but the vector-matter coupling may cause neutrino decay to the exotic vector component.

Using relation (\ref{eqn23}), the phase difference of two mass eigenstates is then given by
\begin{eqnarray}\label{eqn26}
\Phi_{ij}(z) = \int_{z_0}^z \Delta m_{ij}^2 \left[\frac{B^2(\text{z})}{2(E_\nu + g A_0(\text{z}))}\right]d\text{z},
\end{eqnarray}
which is explicitly affected by both non-standard coupling strengths $\beta$ and $g$.
Indeed, we find that the relation we presented for the oscillation phase in curved spacetime is consistent with the one obtained in Refs. \cite{MohseniSadjadi-Yazdani1,MohseniSadjadi-Yazdani2}.
Note that the direction of solar neutrinos propagating to the Earth is assumed to be the $z$-axis.
In other words, we are working at the $\theta_{\text{polar}}=0$ plane, i.e., $z=r$.

Here, considering the wavefunction represented by (\ref{eqn25}), the evolution of the neutrino state can be expressed as follows
\begin{eqnarray}\label{eqn27}
\ket{\upnu(r,t)}_\alpha = \sum_{i,\beta} \mathcal{F}_i (r,t) \upnu_i(r_0,t_0) U_{\alpha i}^* U_{\beta i} \ket{\nu_\beta},
\end{eqnarray}
where $U_{\alpha i}$'s are the elements of the unitary mixing matrix.
In the following, we discuss the effects that occur for the neutrino oscillation probabilities in presence of neutrino decay.
Assume that the flavor of neutrinos produced at the source is $\nu_\alpha$; then, the transition probability of $\alpha \to \beta$ with energy $E_\nu$ propagating in vacuum over a distance $L$ is given by
\begin{eqnarray}\label{eqn28}
P_{\alpha \beta} =\frac{\bigg|\braket{\nu_\beta | \upnu(r,t)}_\alpha\bigg|^2}{\bigg|\braket{\nu_\alpha | \upnu(r_0,t_0)}_\alpha\bigg|^2} =\sum_{i,j} \mathcal{F}_i(r,t) \mathcal{F}_j^*(r,t) U^*_{\alpha i} U_{\beta i} U_{\alpha j} U^*_{\beta j}.
\end{eqnarray}
Defining $c_{ij} \equiv \cos\theta_{ij}$ and $s_{ij} \equiv \sin\theta_{ij}$, the electron-neutrino survival probability is given by
\begin{eqnarray}\label{eqn29}
\begin{split}
& P_{ee} (r) = \mathcal{D}(r) \big[c_{12}^4 c_{13}^4 + s_{12}^4 c_{13}^4  + s_{13}^4  + 2 c_{12}^2 s_{12}^2 c_{13}^4 \cos\left(\Phi_{12}(r)\right) \\&~~~~~~~~~~ + 2c_{12}^2 c_{13}^2 s_{13}^2 \cos\left(\Phi_{13}(r)\right) + 2 s_{12}^2 c_{13}^2 s_{13}^2 \cos\left(\Phi_{23}(r)\right)\big].
\end{split}
\end{eqnarray}
Although more tedious, the probabilities for the other two flavors may be derived in the same way using the approximation $\sin \theta_{13} \ll 1$:
\begin{eqnarray}\label{eqn30}
P_{e\mu}(r) = \mathcal{D}(r) c^2_{23} \sin^2(2 \theta_{12}) \sin^2\left(\frac{\Phi_{12}(r)}{2}\right),
\end{eqnarray}
and
\begin{eqnarray}\label{eqn31}
P_{e\tau} (r) = \mathcal{D}(r) s^2_{23} \sin^2(2 \theta_{12}) \sin^2\left(\frac{\Phi_{12}(r)}{2}\right).
\end{eqnarray}
A typical approximation that successfully describes many experiments with good accuracy is the two-flavor case.
In fact, in the limit where $\sin\theta_{13} \to 0$, compatible with data \cite{Esteban}, oscillations fall into a two-flavor regime.
By substituting from Eq.(\ref{eqn28}), we have the following survival and transition probabilities for the two-flavor case:
\begin{eqnarray}\label{eqn32}
P_{ee}(r) = \mathcal{D}(r) \left[1 - \sin^2 (2\theta) \sin^2\left(\frac{\Phi_{12}(r)}{2}\right)\right],
\end{eqnarray}
and
\begin{eqnarray}\label{eqn33}
P_{e\mu}(r) = \mathcal{D}(r) \sin^2 (2 \theta) \sin^2\left(\frac{\Phi_{12}(r)}{2}\right).
\end{eqnarray}
The fact that the total probability might not be equal to unity is surprising.
It is a direct result of calculating neutrino oscillation probabilities in the presence of neutrino decay.
This indicates that the probability sum $P_{ee} + P_{e\mu} + P_{e\tau}$ is close to unity so that the deviation from unity does not depend on the neutrino energy, but depends on the behavior of the vector field and its coupling to the matter component.

\section{MSW effect and vector-neutrino NSI}\label{sec4}
The MSW flavor transformation is an explicit formulation of the SM and neutrino transition inside matter.
Non-standard interactions with neutrinos and matter components that influence neutrino states can change the $\nu_e$ survival probability prediction.
We include all interactions in the Hamiltonian to see how the coupling to vector boson affects the neutrino flavor change.
For the $2\nu$ case, we then have
\begin{eqnarray}\label{eqn34}
\mathcal{H} = \frac{1}{4E_\nu} \left[\Delta \widehat{m}^2
\begin{pmatrix}
-\cos 2\theta & \sin 2 \theta \\
\sin 2\theta & \cos 2\theta
\end{pmatrix}
+ \mathcal{A}
\begin{pmatrix}
1 & 0 \\
0 & -1
\end{pmatrix}
\right]
,
\end{eqnarray}
where
\begin{eqnarray}\label{eqn35}
\Delta \widehat{m}^2(r) = \left[\frac{B^2(r)}{1 + \frac{g A_0(r)}{E_\nu}}\right] \Delta m^2,
\end{eqnarray}
and $\mathcal{A} = 2\sqrt{2} G_F n_e E_\nu$ is from the effective matter potential, describing the standard interaction of electron-neutrinos with left-handed electrons through the exchange of $W$ bosons.
Using this Hamiltonian in the Schr$\ddot{o}$dinger-like equation
\begin{eqnarray}\label{eqn36}
i \frac{d}{dL}
\begin{pmatrix}
\nu_e\\
\nu_\mu
\end{pmatrix}
= \mathcal{H}
\begin{pmatrix}
\nu_e\\
\nu_\mu
\end{pmatrix}
\end{eqnarray}
for ultra-relativistic neutrinos propagating the distance $L$, one can easily derive the two-flavor neutrino survival probability $<P_{ee}>$.
Henceforth, the critical point is to diagonalize the effective Hamiltonian Eq.(\ref{eqn34}) and to derive the explicit expressions for the effective oscillation parameters.
The solar neutrino parameters of this model correspond to the LMA-MSW solution.
These are related to the vacuum mixing parameters $(\theta, \Delta m^2)$ by
\begin{eqnarray}\label{eqn37}
\sin 2 \theta_{\mathcal{M}} = \frac{\Delta \widehat{m}^{2} \sin 2 \theta}{\Delta m^2_{\mathcal{M}}},
\end{eqnarray}
and
\begin{eqnarray}\label{eqn38}
\Delta m^2_{\mathcal{M}} = \sqrt{ \left[\Delta \widehat{m}^{2} \cos 2\theta - \mathcal{A}\right]^2 +  \left[\Delta \widehat{m}^{2} \sin 2\theta\right]^2 }.
\end{eqnarray}

We have obtained the vacuum limits for the neutrino oscillation probabilities with neutrino decay in the previous section.
In the matter regime, the damping contribution can be derived in the same way as the vacuum case, starting from the following neutrino state for neutrinos produced at the Sun's core and detected on the Earth:
\begin{eqnarray}\label{eqn39}
\ket{\upnu(t,r)}_\alpha = \sum_{i,\beta} \mathcal{F}_i (t,r) \upnu_i(t_0,r_0) U_{\beta i}(\theta,r) U_{\alpha i}^*(\theta_M , r_0) \ket{\nu_\beta}.
\end{eqnarray}
Here, $U_{\alpha i}(\theta_M,r_0)$ and $U_{\beta i}(\theta,r)$ are the mixing matrix elements in the matter at the production point and in the vacuum at the detection point.
Then,
\begin{eqnarray}\label{eqn40}
<P_{ee}> = \mathcal{D}(r) \left[\cos^2\theta \cos^2\theta_{M} + \sin^2\theta \sin^2\theta_{M}\right],
\end{eqnarray}
and
\begin{eqnarray}\label{eqn41}
<P_{e\mu}> = \mathcal{D}(r) \left[\sin^2\theta \cos^2\theta_{M} + \cos^2\theta \sin^2\theta_{M}\right]
\end{eqnarray}
give the damped probabilities on the Earth, where $\theta_{M}$ is the effective mixing angle inside the Sun.
Due to the considerable distance between the Sun and Earth, we could take the averaged oscillation probabilities.
We have defined a general damping signature (\ref{eqn24}) to be an effect that alters the neutrino flavor conversion probabilities to the above forms.
So, the total probability
\begin{eqnarray}\label{eqn42}
<P_{\text{tot.}}> = <P_{ee}> + <P_{e\mu}> \leq 1,
\end{eqnarray}
may not be conserved in such cases.
The equality holds if and only if the $\mathcal{D}$-factor is set to zero.
Expressed in terms of the damping factor, the quantity
\begin{eqnarray}\label{eqn43}
\delta P_{e \to A} = 1- \mathcal{D}(r)
\end{eqnarray}
determines the deviation of the total probability from unity, defined as the ``loss probability''.
The most convincing explanation of transitions among various neutrino flavor eigenstates is probably the neutrino oscillation; however, all the probabilities above also suffer explicitly from damping treatment, indicating an extra deficiency in the neutrino flux on Earth.
This deficiency can be considered an outcome of neutrino decay to vector bosons.
Therefore, we collect these non-standard interactions in a framework together with neutrino oscillations, which are assumed to be the leading order effect and can be constrained by current \cite{SNO-Decay} and future experiments \cite{DUNE,JUNO}.
The present model describes an invisible decay process, which refers to a scenario where the decay products, e.g., a new exotic particle like a vector field, do not interact with the detector's medium. This can also be verified by the screening mechanisms, which suppress vector interactions in dense regions where detections are performed.

\section{Results and discussions}\label{sec5}
Since the discovery of neutrino oscillations has made huge amounts of data available and this phenomenon is now well confirmed by experiments, an understanding of neutrino physics might provide an opportunity to look for a close relationship between neutrinos and dark energy and to put constraints on the dark energy model parameters.
Several auxiliary processes caused by non-standard couplings are interesting topics to explore the dark energy further, as the dominant component of the Universe, describing the present positive cosmic acceleration.
A possibility explaining the solar neutrino problem behind neutrino oscillation might be invisible neutrino decay to the scalar or vector fields, which was first proposed by \cite{Bahcall-decay}.
Experimentally, upcoming long- and short-baseline neutrino observations like DUNE \cite{DUNE} and JUNO \cite{JUNO} will be able to find new signatures of these non-standard couplings.
In the presence of NSI, the evolution of neutrinos is based on the modified Hamiltonian, which encompasses contributions from vacuum, matter, and non-standard interactions.
It is also well known that the propagation of neutrinos in curved spacetime may affect the phase of oscillations.
In the present work, we have studied neutrino interaction with a light vector boson by considering an additional conformal coupling that allows a screening mechanism based on the $\mathbb{Z}_2$ symmetry breaking.
Neutrino propagation inside and outside a spherical object like the Sun can be affected via this NSI, such that both mass and energy terms of the neutrinos are modified, contrary to the scalar field case (see Refs. \cite{Ge-Parke} and \cite{Smirnov-WP}).

\begin{figure}[H]
	\centering
	\includegraphics[scale=0.5]{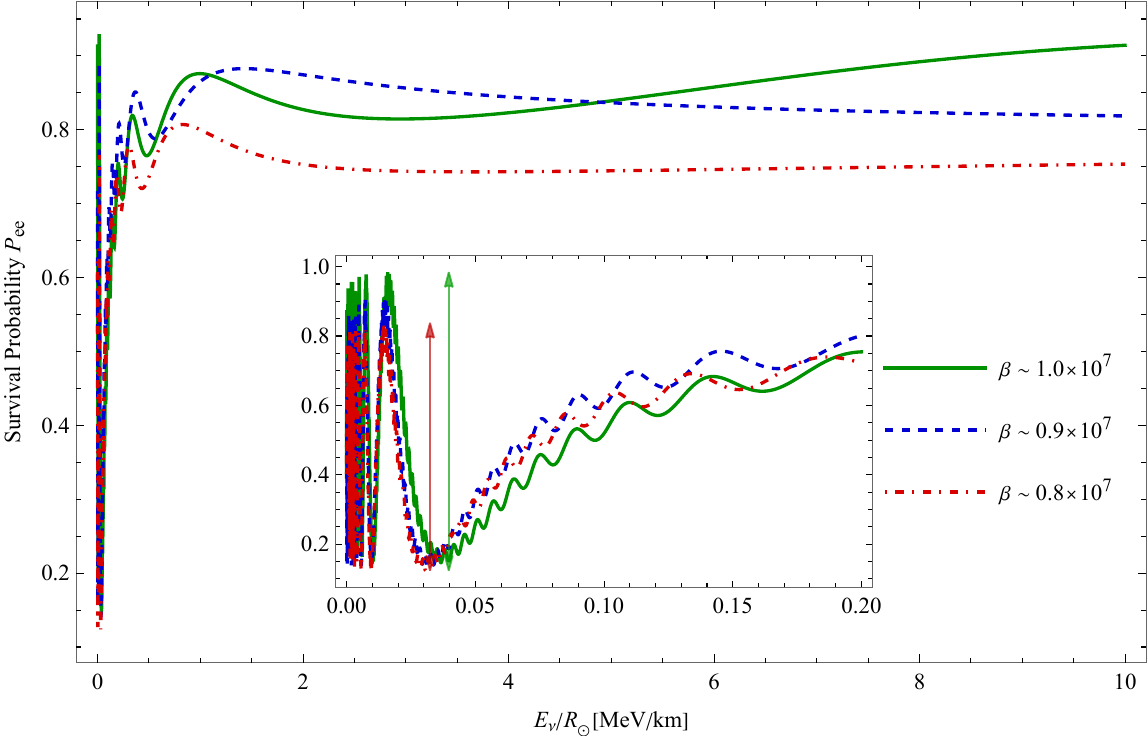}
	\caption{\footnotesize{
			Different curves of $P_{ee}$ in the presence of vector fields for the 3$\nu$ neutrino oscillations.
			This figure has been depicted for the survival probability in terms of (neutrino energy/solar radius).
			Different curves describe $P_{ee}$ corresponding to various values of the vector-matter coupling $\beta$.
			More damping signatures by decreasing matter-neutrino coupling $\beta$ is a result of the vector field-dependent neutrino wavefunction, and consequently, the exponential damping factor $\mathcal{D}$.
		We note that this figure is plotted for $m_A =10^{-27} \text{eV}$ and the vector field-neutrino coupling $g \sim 10^{-9}$.}}
	\label{fig4}
\end{figure}

\begin{figure}[H]
	\centering
	\includegraphics[scale=0.5]{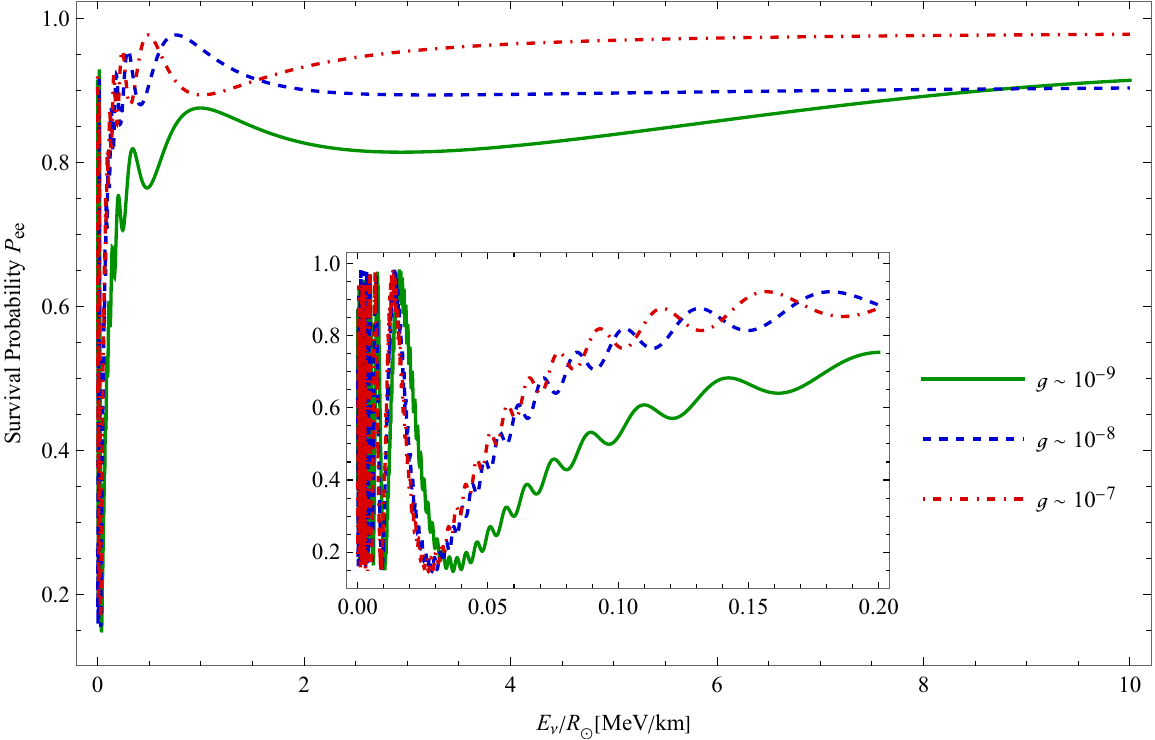}
	\caption{\footnotesize{In this figure, we have depicted the electron-neutrino survival probability in terms of ratio $E_\nu/R_{\odot}$, for $\beta = 1.0 \times 10^7$ and $m_A = 10^{-27} \text{eV}$.
			Different curves describe the $P_{ee}$ for various values of the vector-neutrino coupling parameter $g$.
	Obviously, the constant $g$ does not vary the amplitude of the survival probability, but changes the phase of oscillations.
	}}
	\label{fig5}
\end{figure}

In the following, we numerically illustrate some of the effects of NSI on neutrino flavor conversion.
For this purpose, a wide range of the parameter space is given in Ref.\cite{BeltranJimenez} in which the symmetry is broken outside the Sun and restored inside so that the weak equivalence principle violations and E$\ddot{o}$t-Wash-like experiments can be easily satisfied.
So, we employ the numerical values $\beta \geq 10^6$ and $m_A \gtrsim 10^{-28}$ eV for the coupling strength and vector field's mass, respectively (see the discussion around figure \ref{fig1}).
We have also picked $\tan^2 \theta_{12}=0.41$, $\sin^2 2\theta_{23} \simeq 0.99$, $\sin^22\theta_{13} \simeq 0.09$,
\[ \Delta m_{21}^2=7.4 \times 10^{-5} \text{eV}^2, \] and \[|\Delta m_{23}^2|=2.5 \times 10^{-3} \text{eV}^2\]
for neutrino's mass and mixing parameters \cite{Esteban}.
In Fig.\ref{fig4}, we plot the vacuum survival probability $P_{ee}$ as a function of (neutrino energy/solar radius) for the three flavors to show the effects of the vector conformal coupling on neutrino oscillation.
The resulting profile for the survival probability is sensitive to the vector-matter coupling strength $\beta$ such that this probability will be damped when $\beta$ decreases, contrary to the scalar field case as shown in Ref.\cite{MohseniSadjadi-Yazdani3}, in which the survival probability experienced more damping effects for increasing scalar-matter coupling $\beta$.
The difference in the $\nu_e$ survival probability curves caused by NSI is a mixed situation of an amplitude decrease and a phase shift, which can be regarded as a unique signature for various values of $\beta$, as shown in figure \ref{fig4}.
As illustrated by arrows in the inset of this figure, we find out that the decrease in the amplitude of the much slower oscillations (governed by mass splitting $\Delta m^2_{21}$) is more significant than their fast oscillations (governed by the mass splittings $\Delta m^2_{31}$ and $\Delta m^2_{32}$).
Therefore, vector-matter non-standard damping effects mainly show themselves through amplitude-decreasing effects in the flux of solar electron-neutrinos.
As mentioned, we take two separate vector couplings $\beta$ and $g$ to respectively matter components and neutrino in this model.
Now, let us illustrate the (energy/solar radius)-dependence of the electron-neutrino survival probability in Fig.\ref{fig5} for several vector-neutrino coupling parameters $g$.
From this figure, we clearly see that the vector-neutrino coupling parameter $g$ does not lead to a damping behavior in amplitude of the $\nu_e$ survival probability, as all the curves will tend to a single value at very large $E_\nu/R_\odot$.
So, the only part of vacuum survival probability affected by the vector-neutrino coupling $g$ is the phase of oscillations.
As can be seen from the inset, the curves corresponding to $g \sim 10^{-8}$ and $10^{-7}$ are obviously in opposite phases for fast oscillations.

\begin{figure}[H]
	\centering
	\includegraphics[scale=0.5]{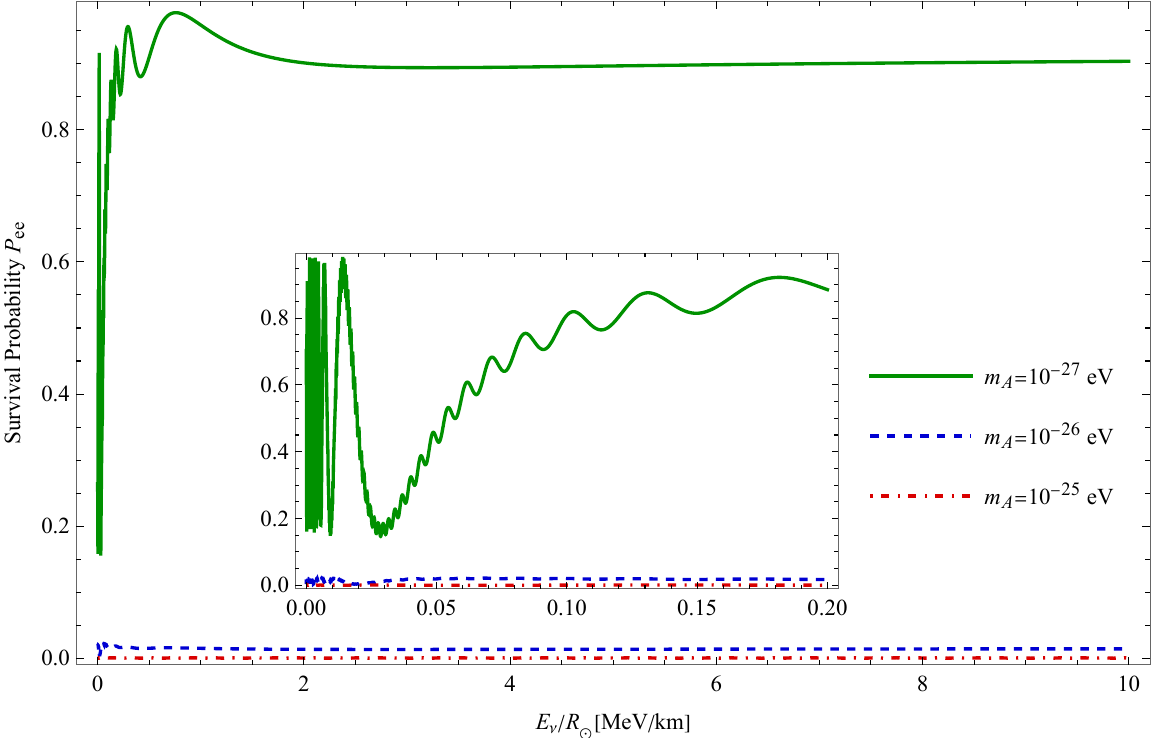}
	\caption{\footnotesize{In this figure, we have depicted the electron-neutrino survival probability in terms of ratio $E_\nu/R_\odot$, for $g =10^{-8}$ and
			$\beta = 1.0 \times 10^7$.
			Different curves describe the $P_{ee}$ for various values of the mass $m_A$.
			Increasing $m_A$ results in totally damped oscillations. According to this figure, $m_A \leq 10^{-27} \text{eV}$ might be reasonable values for the vector field's mass.}}
	\label{fig6}
\end{figure}

\begin{figure}[H]
	\centering
	\includegraphics[scale=0.52]{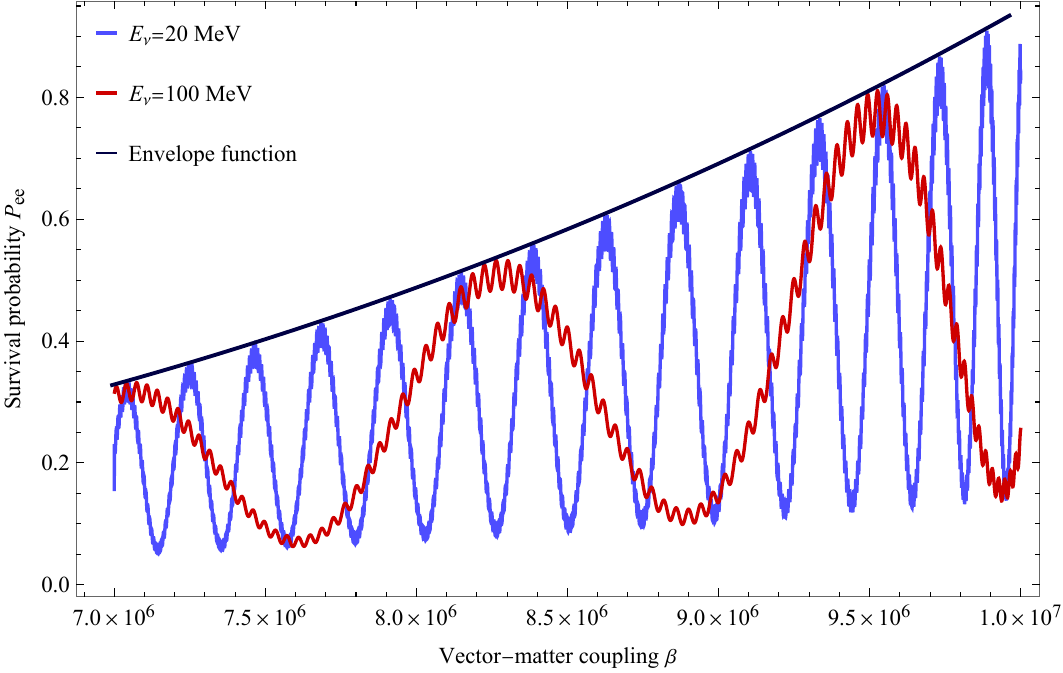}
	\caption{\footnotesize{The behavior of the probability $P_{ee}$ versus the vector field-matter coupling $0.7 \times 10^7 \lesssim\beta \lesssim 1.0 \times 10^7$ for $^8$B solar ($E_\nu = 20~\text{MeV}$) and higher-energy ($E_\nu = 100~\text{MeV}$) neutrinos.
			The envelope function shows how $P_{ee}$ gradually arises when $\beta$ increases.}}
	\label{fig7}
\end{figure}

\begin{figure}[H]
	\centering
	\includegraphics[scale=0.55]{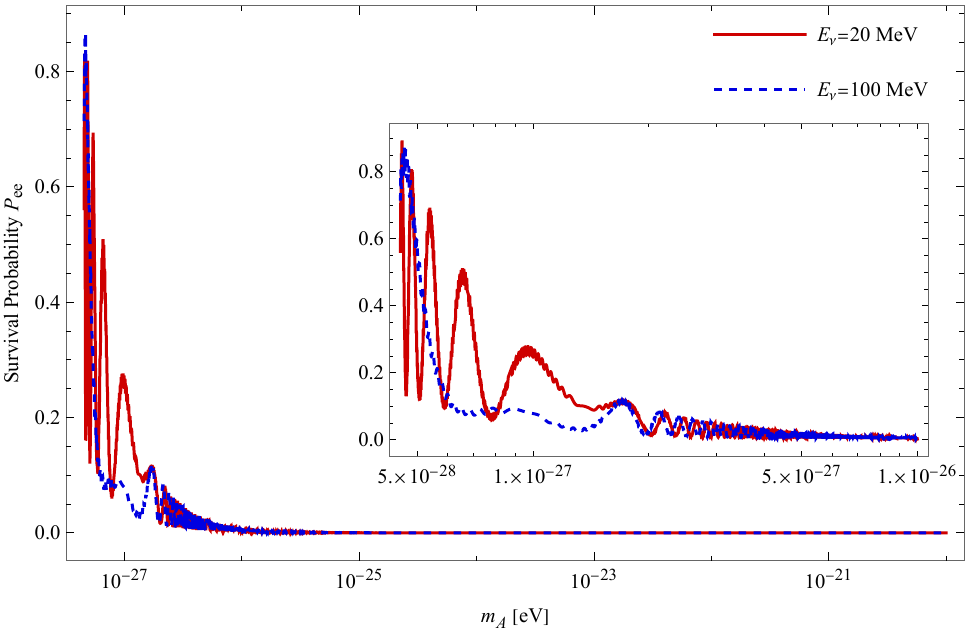}
	\caption{\footnotesize{
From this figure, it becomes evident that for $m_A \geq 10^{-26} \text{eV}$, the survival probability is dramatically damped by the presence of vector NSI for both neutrino energies $E_\nu = 20~\text{MeV}$ and $100~\text{MeV}$.
But the damping of $P_{ee}$ for higher energies is faster than the one of lower energy.
Reasonable values of $P_{ee}$ correspond to $m_A \leq 10^{-27} \text{eV}$.
}}
	\label{fig8}
\end{figure}

In addition, we have plotted the survival probability in terms of $E_\nu/R_\odot$ for different values of another relevant parameter, i.e., the vector field's mass $m_A$ in Fig.\ref{fig6}.
As can be seen, the probability amplitudes of both slow and fast oscillations are suppressed for $m_A \geq 10^{-26} \text{eV}$ such that it has intersection with none of the experimental values for survival probability of $^8$B neutrinos ($E_\nu \simeq 10 \text{MeV}$) measured by SNO ($P^{\text{SNO}}_{ee} = 0.340 \pm 0.023$) \cite{Bellerive} and Borexino ($P^{\text{Borexino}}_{ee} = 0.350 \pm 0.090$) \cite{Agostini}.
As an example, $P_{ee} \simeq 0.346$ for $m_A = 10^{-27} \text{eV}$, while $P_{ee} \simeq 0.018$ for $m_A = 10^{-26} \text{eV}$, which means that the solar electron-neutrinos do not oscillate for $m_A \geq 10^{-26} \text{eV}$.
This inconsistency in survival probability for larger values of $m_A$ can be addressed in discussions about the damping factor (\ref{eqn24}) affected indirectly by $m_A$, and constrains $m_A$ to the values less than $10^{-27} \text{eV}$.
The electron-neutrino survival probability behavior may also be shown as a function of vector-matter couplings $\beta$ for two various values of neutrino energies in Fig.\ref{fig7}.
As previously shown in Fig. \ref{fig4}, this figure confirms less damping effects for increasing $\beta$ (or equivalently, more damping effects for decreasing $\beta$).
Also, the vector-matter coupling $\beta$ might clearly affect the oscillation phase such that $P_{ee}$ has rapid oscillations for increasing $\beta$.
As mentioned, this figure has also been plotted for $^8$B solar neutrinos (with $E_\nu = 20~\text{MeV}$) and higher-energy neutrinos (with $E_\nu = 100~\text{MeV}$) such that higher-energy neutrinos illustrate slower oscillations.
We have already discussed that there is an interesting case, in which the oscillatory term is damped in the presence of vector-matter conformal coupling.
In Fig. \ref{fig8}, we can investigate the influence of neutrino energy $E_\nu$ on neutrino oscillation + decay in terms of the vector field's mass $m_A$.
As shown, the relative importance of the damping signatures increases as the vector field's mass enhances.
Both fast and slow oscillations are suppressed for $m_A \geq 10^{-26} \text{eV}$ and for both neutrino energy values, specially for $E_\nu = 100~ \text{MeV}$, in which the blue-dashed curve shows a sudden drop in the amplitude of $P_{ee}$.
However, in the range $m_A \leq 10^{-27} \text{eV}$, the red curve, plotted for $E_\nu = 20~\text{MeV}$, illustrates a slower damping in the survival probability, which may have an intersection with the observational data of $^8$B neutrinos (see the explanation around Fig.\ref{fig6}).
Hence, we can expect this result puts a constraint on the mass-scale parameter, i.e., $m_A \leq 10^{-27} \text{eV}$, which is also in good agreement with the results mentioned above.

While neutrino oscillation was the first suggested explanation of the solar neutrino problem, we take the idea that the modified MSW-LMA solution specifies the correct picture and will consider neutrino decay as a sub-dominant effect.
That is, besides the conversion into three known active neutrinos through neutrino oscillations, we consider the so-called ``invisible decay'', e.g., to light vector boson.
Neutrino decay process results in a depletion of the total flux of neutrinos by the damping factor presented in Eq.(\ref{eqn24}).
Now, we try to assess the neutrino loss probability with various model parameters.

\begin{figure}[H]
	\begin{subfigure}{.5\textwidth}
		\centering
		\includegraphics[scale=0.46]{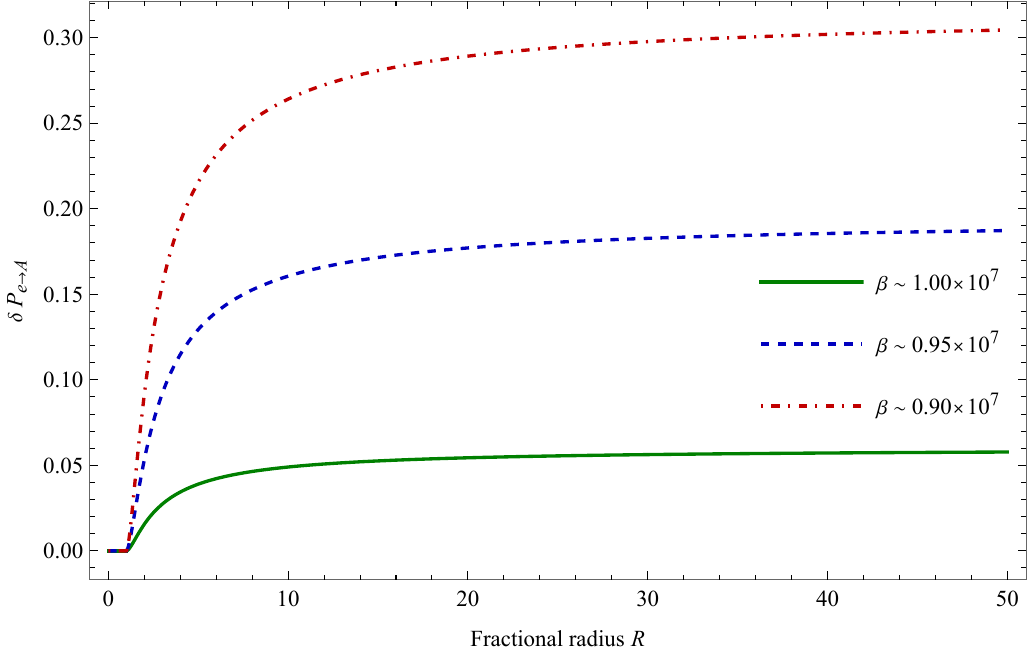}
		\label{fig9-1}
		\caption{\footnotesize{}}
	\end{subfigure}
	\begin{subfigure}{.5\textwidth}
		\centering
		\includegraphics[scale=0.48]{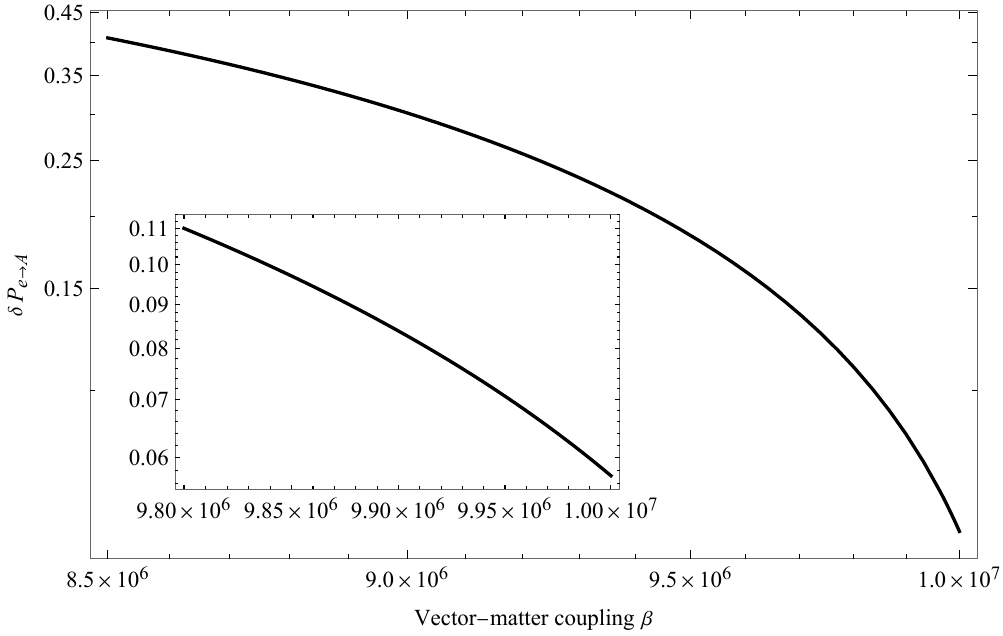}
		\label{fig9-2}
		\caption{\footnotesize{}}
	\end{subfigure}
	\caption{\footnotesize{The loss probability of $^8$B electron-neutrinos on the Earth with respect to the fractional radius $R$ (Panel (a)), and with respect to the coupling parameter $\beta$ (Panel (b)).
			Each probability is calculated numerically, using $\rho_0 \sim 10^{-24} \text{g}.cm^{-3}$ for background matter density.}}
	\label{fig9}
\end{figure}

In the panel (a) of Fig. \ref{fig9}, we display the solar neutrino loss probability $\delta P_{e \to A}$ as a function of fractional distance from the sun's center, for different couplings $\beta$.
At large fractional radii $R$, the height of each curve is set conversely by the coupling parameter $\beta$ (see relation (\ref{eqn43})).
The solid green curve corresponds to the case of $\beta \sim 1.0 \times 10^7$, which asymptotically tends to the $\delta P_{e \to A} \simeq 0.06$ on the Earth.
In the panel (b), we show the corresponding loss probability with respect to the coupling $\beta$, which explicitly explains how the larger coupling $\beta$ leads to a lower loss probability.
Inset of this panel shows the loss probability near $\beta = 1.0 \times 10^7$.

The constraints on different model parameters might be derived from the analysis of neutrino flavor conversion data, which may be held true for a combination of SM and NSI and, consequently, for the modified MSW effect.
An interesting property of the present model is its $\beta$ and $g$ dependencies, in which we have considered the SNO+SK \cite{Zyla} and Borexino \cite{Agostini} survival probability data.

\begin{figure}[H]
	\centering
	\includegraphics[scale=0.45]{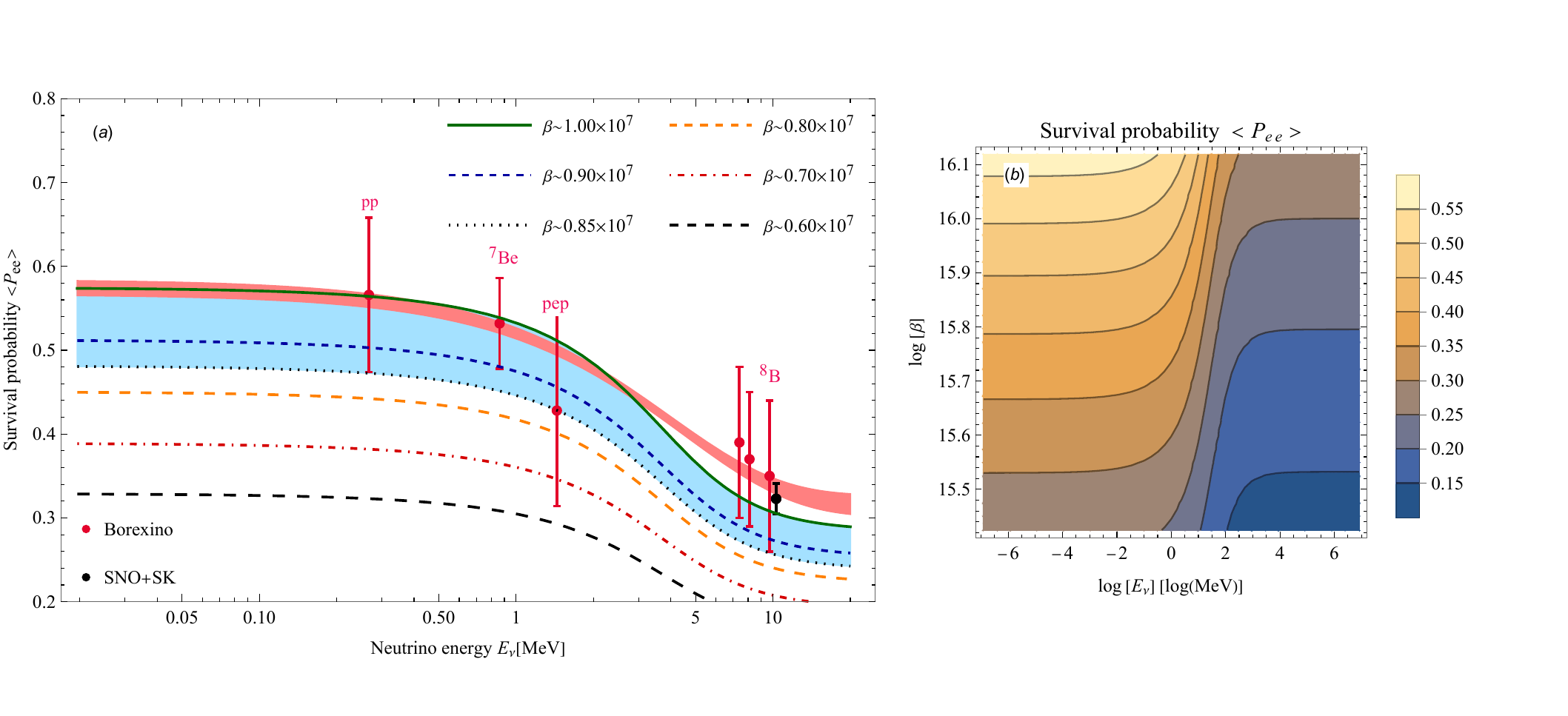}
	\caption{\footnotesize{Panel (a): The survival probability as a function of energy with various LMA-MSW + NSI cases.
		The pink band shows the standard effects.
		We have assumed that $g \sim 10^{-10}$ and $m_A = 10^{-27} \text{eV}$.
		Panel (b):The electron-neutrino survival probability $<P_{ee}>$ as a function of coupling parameter $\beta$ and neutrino energy $E_\nu$. The colored regions correspond to the numerical results (darker regions for lower probabilities).
	For $0.85 \times 10^7 \leq \beta \leq 1.00 \times 10^7$, the curves are mostly within the error bars of Borexino data (the blue region shown in the plot).}}
	\label{fig10}
\end{figure}

Figure \ref{fig10} illustrates the analysis of survival probability for six various values of the vector-matter coupling parameters within the solar neutrino experiments SNO+SK and Borexino.
High-energy solar neutrinos ($^8$B and $hep$ neutrinos) experience adiabatic resonant conversion within the Sun (with a survival probability of about $0.33$), while the flavor changes of the lower energy solar neutrinos ($pp$, $^7$Be, and $pep$ neutrinos) arise only from vacuum oscillations.
These vacuum oscillations lead to an average survival probability of about $0.55$, as implied by the present Borexino data.
The transition from the matter flavor conversion inside the Sun to the vacuum oscillations lies in $3-4$ MeV.
This makes $^8$B neutrinos the best choice when looking for a transition point within the energy spectrum.
It is in this transition region where the physics BSM might be most considerable, as they have intersections with the curve from standard matter effects (light-pink band).
The light-pink band is the best theoretical prediction of $\nu_e$ survival probability (within $\pm 1\sigma$) according to the standard MSW-LMA solution.

In the panel \ref{fig10}(a), it becomes evident that in the low-energy range, curves with the couplings $\beta < 0.85 \times 10^7$ undergo more damping effects and drop to the constant value $<P_{ee}> \sim 0.33$, which do not agree with the experimental data.
While one might think that the lowest energy neutrinos (i.e., the obtained data of $pp$, $^7$Be and $pep$ neutrinos from Borexino) are the best candidates for showing decay effects, there is an important issue;
the Borexino experiment measures various energy neutrinos, and receives contributions from several unavoidable background sources, basically coming from the radioactive isotopes contaminating the scintillator itself \cite{Agarwalla} or even from cosmic rays hitting the Earth's atmosphere \cite{Kumaran}.
These background effects on electron-neutrino survival probability can disturb the results obtained from physics BSM.
So, in the energy range covered by these experiments, the physics BSM can only be explored for solar neutrino flux dominated by the $^8$B neutrinos.
Therefore, in the high-energy domain, for the vector-matter coupling values $0.85 \times 10^7 \leq \beta \leq 1.00 \times 10^7$, the different curves are in good agreement with observations from $^8$B neutrinos, as they are mostly within the error bars of the experimentally determined values of Borexino, i.e., the values of $\beta$ limited to the blue region shown in the plot.
As a direct result of figure \ref{fig7}, the damping signatures grow when the vector-matter couplings $\beta$ decreases, which can also be explicitly concluded from this figure.
This result is contrary to the scalar dark energy model (refer to \cite{MohseniSadjadi-Yazdani3}), in which the survival probability experienced more damping for larger values of the coupling strengths $\beta$.

In the panel \ref{fig10}(b), the contour plot of the survival probability with respect to the $\beta$-$E_\nu$ is drawn.
This plot clearly demonstrates that large vector-matter couplings $\beta$ are corresponding to the lower values of the $\nu_e$ survival probability, which obviously confirms considerable damping signatures for lower values of $\beta$.

Also, we note that the optimal flavor mixing occurs whenever $\sin^2 2\theta_\mathcal{M} =1$, resulting in the resonance energy, i.e.,
\begin{eqnarray}\label{eqn44}
E_{\nu}^{\text{res.}} = \frac{\Delta m^{2} B^2(r_{\text{res.}}) \cos 2 \theta}{2\sqrt{2}G_F n_{e}(r_\text{res.})} - g A_0(r_{\text{res.}}).
\end{eqnarray}
Since the resonance energy is explicitly dependent on the vector-matter coupling strengths $\beta$ and vector-neutrino coupling coefficient $g$, it is expected that there would be a shift in the resonance energy for various coupling strengths.
From this figure, it is evident that the resonance energy grows when we increase the coupling $\beta$.

\begin{figure}[H]
	\centering
	\includegraphics[scale=0.45]{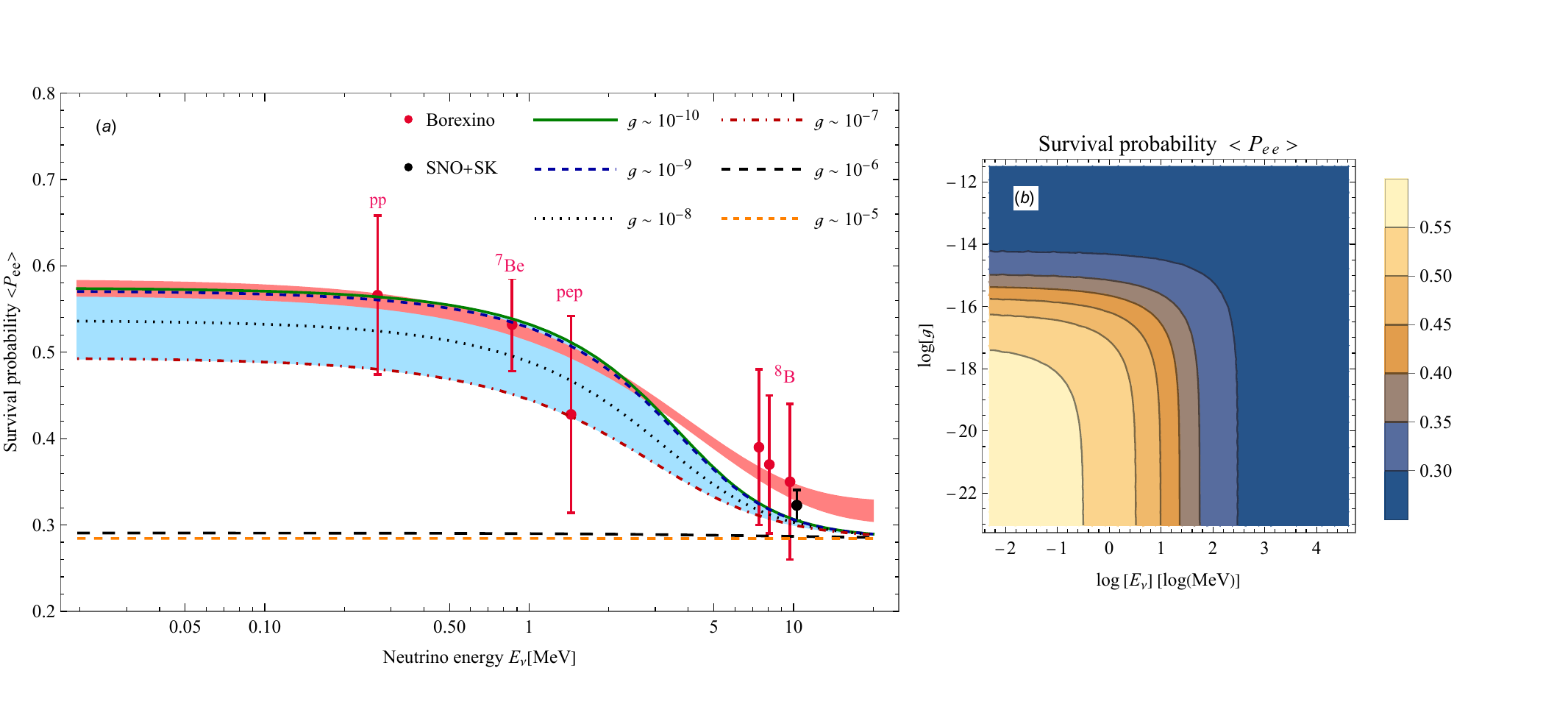}
	\caption{\footnotesize{Panel (a): The solar neutrino survival probabilities with the vector NSI, together with the Borexino measurement \cite{Agostini} of the pp, $^7$Be, pep, and $^8$B fluxes.
		The black point also represents the SNO + SK $^8$B data \cite{Zyla}.
		This figure is plotted for $\beta \sim 1.0 \times 10^7$ and $m_A = 10^{-27} \text{eV}$.
Panel (b): We present the contour plot of neutrino survival probability $<P_{ee}>$ for different values of the coupling parameter $g$ and neutrino energy.
	For $10^{-10} \leq g \leq 10^{-7}$, the curves are mostly within the error bars of Borexino data (the blue region in the panel (a)).}}
	\label{fig11}
\end{figure}

We have also re-examined the modified MSW effects in our theoretical framework for different vector-neutrino couplings $g$ and suggested an allowed range for this parameter corresponding to the observational data \cite{Agostini}.
In the panel \ref{fig11}(a), different curves in energy regions of about $20$ MeV are the least influenced part by the presence of NSI across all six various $g$-varying configurations.
Additionally, it is noteworthy that, in the low energy regions (i.e., for $E_\nu \leq 5$ MeV), the impact of NSI mostly deviates further for $g \geq 10^{-6}$ such that the survival probability curves drop to a constant value of around $0.33$, whereas NSI is most considerable for $10^{-10} \leq g \leq 10^{-7}$ such that the curves are mainly within the error bars of experimentally obtained values of Borexino at all energies, as shown with the blue region in the panel (a).
The vector-neutrino coupling $g$ might affect the amplitude of $<P_{ee}>$.
This observation can be attributed to the fact that the coupling coefficient $g$ contributes to the modified mixing parameters (\ref{eqn37}) and (\ref{eqn38}) such that $\cos 2\theta < \frac{\mathcal{A}}{\Delta \widehat{m}^2}$.
Also, the non-standard coupling $g$ may affect the resonance energy such that larger couplings $g$ result in lower values of $E^{\text{res.}}_{\nu}$.

The contour plot of the survival probability with respect to the coupling parameter $g$ and neutrino energy has been drawn in Fig. \ref{fig11}(b), providing a clearer continuous representation of the $g$-$E_\nu$ dependence of the survival probability.

\begin{figure}[H]
	\begin{subfigure}{.5\textwidth}
		\centering
		\includegraphics[scale=0.52]{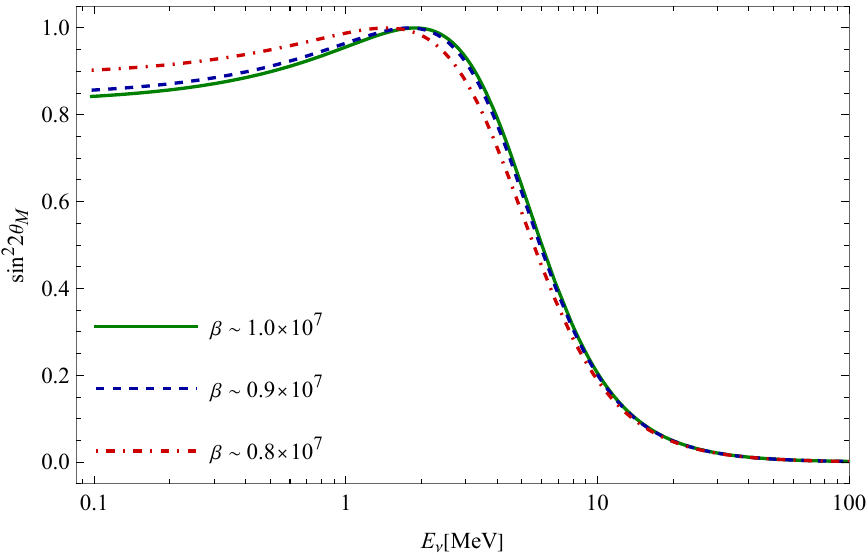}
		\label{fig12-1}
		\caption{}
	\end{subfigure}
	\begin{subfigure}{.5\textwidth}
		\centering
		\includegraphics[scale=0.52]{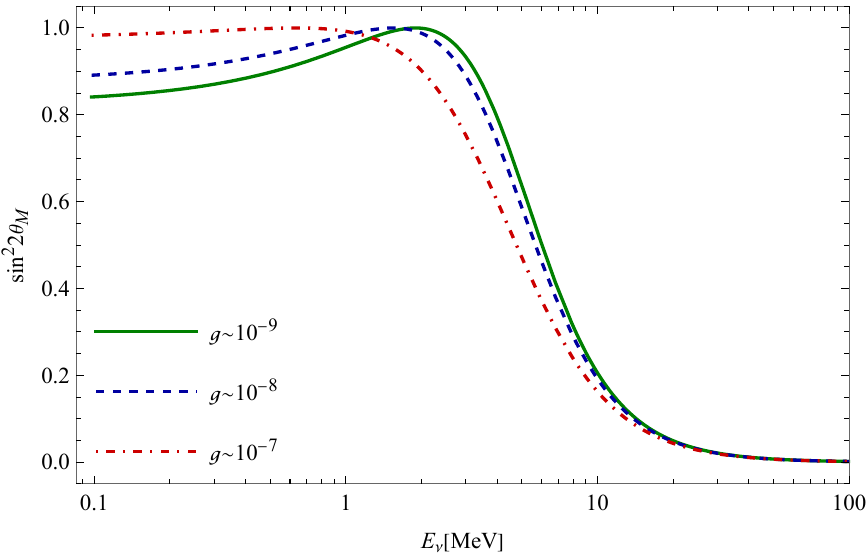}
		\label{fig12-2}
		\caption{}
	\end{subfigure}
	\caption{\footnotesize{$\sin^2 2\theta_M$ as a function of neutrino energy $E_\nu$ for vector-matter coupling $\beta$ (Panel (a)), and for vector-neutrino coupling $g$ (Panel (b)).}}
	\label{fig12}
\end{figure}

According to Eq.(\ref{eqn44}), the effective mixing parameter $\theta_M$ in matter depends not only on the electron number density and neutrino energy, but also is a function of couplings $\beta$ and $g$.
In Fig. \ref{fig12}, we show the change of $\sin^2 2\theta_M$ in terms of neutrino energy, taking various vector coupling strengths $\beta$ and $g$.
The panel (a) illustrates the $\theta_M$-dependence on the vector-matter coupling $\beta$, in which different curves correspond to the $\beta = 1.0 \times 10^7, 0.9\times 10^7$, and $0.8\times 10^7$.
At low energies (for $E_\nu \sim 2$ MeV), $\sin^2 2\theta_M \to 1$ shows the maximal mixing at the peaks, which are shifted to the higher energies when $\beta$ grows.
In the panel (b), however, we show the behavior of modified mixing angle inside matter for various vector-neutrino couplings $g \in \{10^{-9}, 10^{-8}, 10^{-7}\}$.
The maximum values are shifted to the higher energies by decreasing $g$.
In both cases, as the neutrino energy increases, the mixing inside matter goes to zero.
So, $\sin^2 2\theta_M$ becomes negligible, and thus the transition probability becomes low at $E_\nu > 100~\text{MeV}$.

\section{Conclusion}\label{sec6}
A Symmetron-like vector screening model, consisting of a massive vector field that is conformally coupled to matter, was introduced.
We have obtained the vector field solution by expanding the field around its background value, i.e.,  $A_0 = A_0^{\text{min}} + \delta A_0$ (see Eqs.(\ref{eqn10}) and (\ref{eqn11})).
The solution to the Dirac equation was derived and employed to study neutrino flavor oscillation.
The NSI modifies the solar neutrino transition probabilities via both vacuum oscillations and the MSW effect as the neutrinos propagate through dense solar matter.
We obtained the effective Hamiltonian induced by both the matter electrons and the vector component (refer to Eq.(\ref{eqn34})).
From the damping factor $\mathcal{D}$, we concluded that the sum of various probabilities is assumed to be not equal to unity, meaning that there is an invisible decay possibility along neutrinos' journey (see Eq.(\ref{eqn43})).

In this context, the possible presence of non-standard interactions (NSI) of neutrinos and matter components with a vector dark energy was discussed from a phenomenological perspective, outlying  the current status of dark energy cosmology research.
To date, there is no local experimental evidence or indication of the presence of dark energy. However, screening mechanisms are necessary to suppress the vector field on small scales and produce observable signals on cosmological scales.
Proposed near-future neutrino experiments like DUNE \cite{DUNE} and JUNO \cite{JUNO} are expected to either indicate the existence of NSIs or significantly improve the bounds on NSIs. While our ability to detect neutrinos is continuously improving, the theoretical understanding of neutrino properties and their various beyond-standard-model (BSM) interactions remains an open question.
Some earlier studies concluded that vector non-standard interactions (NSIs) affect neutrino flavor change similarly to scalar NSIs \cite{Ge-Smirnov}. However, our findings suggest that vector NSIs might yield different results compared to scalar NSIs:

$\bullet$ The fact that the vector and scalar NSIs appear differently in the Dirac Lagrangian leads to totally different effects on the neutrino evolution.
Vector interactions impact directly not only the neutrino mass ($m^\prime \bar{\upnu}^\prime \upnu^\prime$ with $m^\prime = m B$ and $\upnu^\prime = B^{\frac{3}{2}} \upnu$) but also the neutrino energy (through the interacting term $g \bar{\upnu}^\prime \gamma^\mu A_\mu \upnu^\prime$, resulting in $E_\nu \to E_\nu + gA_0$), see the Dirac equation (\ref{eqn20}).
Therefore, vector NSIs have a significant effect at all energy ranges, even for low-energy neutrinos (see Figs. \ref{fig10} and \ref{fig11}), specifically when the non-standard vector-neutrino coupling term $g A_0$ is of the same order as neutrino energy (see the discussion around relation (\ref{eqn23})).
For the scalar field case, however, since the scalar interaction cannot be converted into a vector current, which is verifiable by explicit derivations \cite{MohseniSadjadi-Yazdani3}, the light scalar field does not give rise to a modification in matter potential and only contributes as a correction to the neutrino mass through a Yukawa NSI term.

$\bullet$ Furthermore, a comparison of scalar NSI \cite{MohseniSadjadi-Yazdani3} reveals that the scalar-matter coupling strength is limited to $\beta \sim 100$.
That is, for the coupling strength of order $\beta >100$, the scalar NSI case remains excluded, whereas the vector NSI case has viable parameter spaces of order $0.85\times 10^7 \leq \beta\leq 1.00\times 10^7$ and $10^{-10}\leq g \leq 10^{-7}$, as has been discussed in the previous section.

Using neutrino flavor conversion probes provides a more precise assessment of dark energy properties, such as its mass range.
Current data from Borexino \cite{Agostini} and SNO+SK \cite{Zyla} can also constrain the vector field's mass to  $10^{-28} \text{eV}\lesssim m_A \lesssim 10^{-27} \text{eV}$ in the present model, where the vector field represents a fraction of dark energy, rather than its entirety.
As explored in Ref. \cite{Fujita}, the mass region of an exotic field might be $10^{-42} \text{eV} \lesssim m_A \lesssim 10^{-25.5} \text{eV}$, in which the field can act as either dark energy or a sub-dominant component of dark matter, depending on its mass \cite{Frieman,Kim1,Kim2,Choi,Nomura,Kim3,Hall,Kim4,Chatzistavrakidis,Kim5,Kang}.
For $m_A \lesssim H_0$, the energy density of the field can account for all of dark energy. In this mass $\Omega_A$ can be as large as the dark energy density parameter $\Omega_A \sim \Omega_\Lambda = 0.69$.
However, if $m_A$ is larger than $H_0$, the energy density of the field is not the main source for dark energy.
This means that for the mass range $10^{-32} \text{eV} \lesssim m_A \lesssim 10^{-25.5} \text{eV}$, the observations of CMB and large-scale structures put constraints on the relative density $\Omega_A \leq 0.006 h^{-2}$, where the dimensionless constant $h$ is taken to be $0.677$ \cite{Hlozek}.
The mass range we determined for the vector field is fully incorporated in this case.
The field is primarily responsible for dark matter in the much higher mass region, and its clustering may play an important role in the axion-like particle (ALP) models mentioned by \cite{Fujita}.

\end{document}